\documentclass[12pt,a4paper]{article}
\begin{document}
\sloppy
\title{Advances on String Theory \\ in Curved Space Times}
\author{N.G. SANCHEZ\footnote{Norma.Sanchez@obspm.fr}\\
Observatoire de Paris, LERMA\\61,avenue de l'Observatoire\\ 75014 Paris,
 FRANCE}
\date{}
\maketitle

\noindent
String Quantum Gravity is motivated and introduced.\\
Advances in the study of the classical and quantum string
dynamics in curved spacetime are reported :

\noindent
1 - New Classes of Exact Multistring Solutions in curved Spacetimes.\\
2 - Mass Spectrum of Strings in Curved Spacetimes.\\
3 - The Effect of a Cosmological Constant and of Spatial Curvature on\\
    Classical and Quantum Strings.\\
4 - Classical Splitting of Fundamental Strings.\\
5 - The General String Evolution in Constant Curvature Spacetimes.\\
6 - The Conformal Invariance Effects. \\ \\
In 1987, we started a programme \cite{vs1} to study the string dynamics in 
curved
spacetime and its associated physical phenomena. This study revealed new
insights and new physical phenomena with respect to string propagation in 
flat spacetime (and with respect to quantum fields in curved spacetime). 
The results are relevant both for fundamental (quantum) 
strings and for cosmic strings, which behave essentially in a classical way. 
Approximative and exact solving methods have been developed. Classical and 
quantum string dynamics have been investigated in black hole spacetimes,
cosmological backgrounds, cosmic string spacetime, gravitational wave 
backgrounds, supergravity backgrounds (which are necessary for fermionic
strings), and near spacetime singularities. Physical phenomena like 
classical string instability and non oscillatory motion in time, quantum 
particle transmutation, string scattering, string stretching, have been 
found. For the results (1987-1994) see for example our Chalonge Erice Lectures 1989-1994 and references therein. See (http://www.obspm.fr/chalonge), 
and ``String Theory in Curved Space Times'', Ed. N. Sanchez, WSPC, (1998). \\

\section{Multistring Solutions: A new feature for strings in curved spacetime} 
The discovery of the multistring property \cite{cvms} \cite{vms} in the 
propagation of strings
in curved spacetimes is the consequence of several developments : \\ \\
(i) The classical string equations of motion plus the string constraints were
shown to be exactly integrable in D-dimensional De Sitter spacetime and 
equivalent to a sinh-Gordon model with a Hamiltonian unbounded from below \cite{vs2}.
 Generalization of this result including the Cosh-Gordon and Liouville 
equations, for strings and multistrings in constant curvature spacetimes have 
been given in ref \cite{ls5}.\\ \\
(ii) \begin{em}
Exact 
\end{em}
string solutions were systematically found by soliton methods using the linear system associated to the problem (the so-called dressing method in soliton
theory) \cite{cvms}. In particular, exact circular string solutions were found in 
terms of elementary \cite{vms} and elliptic functions \cite{vls1}.\\ \\
(iii) All these solutions describe one string, several strings or even an
infinite number of different and independent strings. A single world-sheet
simultaneously describes 
\begin{em}
many
\end{em}
different strings. This new feature appears as a consequence of the coupling
of the strings with the spacetime geometry. Here, interaction among the 
strings (like splitting and merging) is neglected, the only interaction is
with the curved background. \\
Different types of behaviour appear in the multistring solutions. For some of 
them the energy and proper size are bounded (``{\it stable strings}'') while 
for
many others the energy and proper size blow up for large radius of the 
universe (``{\it unstable strings}''). \\ \\
In all these works, strings are {\it test} objects propagating on the given
fixed backgrounds. The string energy momentum tensor was computed and the 
string equation of state {\it derived} from the string dynamics in cosmological and black hole spacetimes. Strings obey the perfect fluid relation 
\begin{displaymath}
 p=(\gamma-1)\rho 
\end{displaymath} 
with three different behaviours :\\ (i) {\it Unstable} for 
large R, with negative pressure;\\ (ii) {\it Dual} for small R, with positive
pressure, (as radiation);\\ (iii) {\it Stable} for large R, with vanishing 
pressure, (as cold matter). \\ \\
We find the {\it back reaction effect} of these strings on the spacetime \cite{vs3}. 
This is achieved by considering {\it selfconsistently} the strings as matter 
sources for the Einstein equations, as well as for the complete effective 
string equations, for cosmological spacetimes at the classical level. The
selfconsistent solution of the Einstein-Friedman equations for string 
dominated universes exhibits the realistic matter dominated behaviour for 
large times and the radiation dominated behaviour for early times. That is, 
the {\it standard cosmological} evolution is well generated by strings. It 
must be noticed that there is no satisfactory derivation of inflation in the
context of the effective string equations. De Sitter universe {\it does not} 
emerge 
as solution of the effective string equations. The effective string action 
(whatever be the dilaton, its potential and the central charge term) is not 
the appropriate framework in which to address the question of string driven 
inflation. \\ \\
More recently, \cite{ls1}, new classes of exact {\it \bf multistring} solutions were found. The multistring solutions were classified and their physical 
properties described.\\ In Anti de Sitter spacetime, the solutions describe an 
{\it \bf infinity} number of infinitely long stationary strings of equal energy but different pressures. In De Sitter spacetime, outside the horizon, 
they describe infinitely many {\it \bf dynamical} strings, infalling non radially, scattering at the horizon and going back to spatial infinity in 
different directions. For special values of the constant of motion, there are 
families of solutions with {\it \bf selected finite} numbers of different and independent strings. The strings appear {\it \bf distributed} in 
{\it \bf packets}, the number of strings in each packet, the number of 
``turns'' or ``festoons'' in each string is precisely determined and solely 
dictated by the dynamics, exactly solved in terms of elliptic functions. \\ \\
In Black hole spacetimes, (without cosmological constant) no multistring 
solutions are found \cite{ls2}. In the Schwarzschild black hole, inside the 
horizon, 
the string infalls, with {\it \bf indefinitely} growing size and energy, into the r=0 singularity and the string motion stops there.\\ In the (2+1)- black hole anti-de Sitter background, the string stops at r=0 with {\it \bf  
finite} length \cite{ls3}; the reason being that the point r=0 is not a strong 
curvature singularity in the (2+1)-black hole anti- de Sitter spacetime. 
Outside the horizon, in this spacetime, the multistring solution describes 
infinitely many, infinitely long open strings. 
\section{The String Mass Spectrum in the presence of Cosmological Constant}
The string mass spectrum in the presence of a cosmological constant (for both 
de Sitter and Anti de Sitter spacetimes) was found \cite{ls3}, \cite{vls2}. 
New features as 
compared to the string spectrum in flat spacetime appear, as a {\it \bf 
fine structure effect} (splitting of levels) at all the states beyond the 
graviton, (in both de Sitter (dS) and AntideSitter (AdS) spacetimes), and the 
{\it \bf absence} of a critical Hagedorn temperature in AdS spacetime 
(the partition function for a gas of strings in AdS spacetime is well defined 
for all temperature).\\ \\ The presence of a cosmological constant reduces 
(although do not totally removes) the degeneracy of states as compared with 
flat Minkowski spacetime. In AdS spacetime, the density of states $\rho$(m) 
grows like $\exp$[($\Lambda$ m)$^\frac{1}{2}$], (while, as is known, in 
Minkowski spacetime, $\rho$(m) grows like the Exponential of the mass m). The 
high mass spectrum changes drastically with respect to flat Minkowski spacetime. The level spacing {\it \bf 
grows} with the eigenvalue of the number operator, N, in AdS spacetime, while 
is approximatively constant (although smaller than in Minkowski spacetime and 
slightly decreasing) in dS spacetime. There is an infinite number of states 
with arbitrarily high mass in AdS space time in dS there is a {\it \bf finite} number of oscillating states only.\\The string mass has been expressed in terms of 
the Casimir operator C = L$_{\mu\nu}$ L$^{\mu\nu}$ of the O(3,1) De Sitter 
group [O(2,2) group in Anti deSitter \cite{vls2}. See also section 3 below.
\section{Spatial Curvature Effects}
The effects of the spatial curvature on the classical and quantum string 
dynamics is studied in ref.\cite{ls4}. The general solution of the circular 
string motion in static Robertson-Walker spacetimes with closed or open 
sections has been found \cite{ls4}. 
This is given closely and completely in terms of elliptic functions.\\
The {\it \bf back reaction effect} of these strings on the spacetime is 
found : the self-consistent solution to the Einstein equations is a spatially 
closed ({\bf K}$>$0) spacetime with a selected value of the curvature index {\bf K}, {\bf K} = ({\bf G}/${\bf \alpha}$')$^
\frac{2}{3}$ (the scale factor is normalized to unity). No self-consistent solutions with {\bf K}$<$0 exist. We semiclassically quantize the circular 
strings and find the mass m in each case. For {\bf K}$>$0, the very 
massive strings, oscillating on the full hypersphere, have 
\begin{displaymath}
{\bf m}^2 \sim {\bf K} {\bf N^2}, ({\bf N} {\bf \epsilon}  
{\bf N}_0 )
\end{displaymath}
 {\it \bf independent} of ${\bf \alpha}$' and the 
level spacing {\it \bf grows} with n, while the strings oscillating on 
one hemisphere (without crossing the equator) have 
\begin{displaymath}
{\bf m}^2 \sim {\bf \alpha}' {\bf N}
\end{displaymath}
 and a {\bf finite} number of states 
N $\sim$ 1/(K${\bf \alpha}$'). For K $<$0, there are infinitely many 
strings states with masses
\begin{displaymath} 
{\bf m} \quad {\bf \log} {\bf m} \sim {\bf N},
\end{displaymath} 
that is the level spacing grows {\it \bf slower} than {\bf N}.\\The stationary string solutions as well as the generic 
string fluctuations around the center of mass are also found and analyzed in 
closed form.
\section{Classical Splitting}
We find exact solutions of the string equations of motion and constraints 
describing the classical splitting of a string into two \cite{vmms}. For 
the same Cauchy data, the strings which split have smaller action that the 
string without splitting. This phenomenon is already present in flat space-
time. The splitting process takes place in real (lorentzian signature 
spacetime).\\The solutions in which the string splits are perfectly natural 
within the classical theory of strings. There is no need of extra interactions, (nor extra terms in the action to produce splitting). The difference with the 
non splitting solutions is on the boundary conditions. \\The mass, energy and momentum carried out by the strings are computed. We show that the splitting 
solution describes a natural decay process of one string of mass {\bf M} 
into two strings with a smaller total mass and some kinetic energy. The 
standard non-splitting solution is contained as a particular case. \\We also 
described 
the splitting of a closed string in the background of a singular gravitational 
plane wave, and showed how the presence of the strong gravitational field 
increases (and amplifies by an overall factor) the negative difference between the action of the splitting and non-splitting solutions.
\section{General String Evolution in Constant Curvature Space-Times}
In ref. \cite{ls5}, we have found that the fundamental quadratic form of the 
classical string propagation in (2+1)-dimensional constant curvature 
spacetimes, solves the sinh-Gordon equation, the cosh-Gordon equation, or the 
Liouville equation. In both de Sitter and anti-de Sitter spacetimes, (as well 
as in the 2+1 black hole anti-de Sitter spacetime), {\it all} three equations 
must be included to cover the generic string dynamics. This is particularly 
enlightening since {\it generic} properties of the string evolution can be thus {\it directly} extracted from the properties of these three equations and 
their associated Hamiltonians or potentials, {\it irrespective of any solution}. \\These results complete and generalize our previous results on this topic 
(until now, only the sinh-Gordon sector in de Sitter spacetime was known). \\
We also construct new classes of multistring solutions, in terms of elliptic 
functions, to all three equations in both de Sitter and anti de Sitter 
spacetimes, which generalize our previous ones. \\These results can be 
straighforwardly generalized to constant curvature spacetimes of arbitrary 
dimension, by replacing the sinh-Gordon equation, the cosh-Gordon equation, and the Liouville equation by their higher dimensional generalizations. \\
Our results indicate the existence of various kinds of dualities relating the 
different sectors and their solutions in de Sitter and anti-de Sitter 
spacetimes : in the sinh-Gordon sector of de Sitter spacetime, small strings 
are dual (that is, under S $\rightarrow$ 1/S, S being the proper string size, they are mapped) to large strings. And, similarly, in the sinh-Gordon sector of 
anti-de Sitter spacetime. Furthermore, in the cosh-Gordon sector, small (large) strings in de Sitter spacetime are dual to large (small) strings in the 
anti-de Sitter spacetime.
\section{Conformal Invariance Effects}
Classical and quantum strings in the conformally invariant background 
corresponding to the $SL(2R)$ WZWN model has been studied in ref \cite{vls3}. 
This background is locally anti-de Sitter spacetime with non-vanishing 
torsion. Conformal invariance is expressed as the torsion being parallelizing; 
and the precise effect of the conformal invariance on the dynamics of both 
circular and generic classical strings has been extracted \cite{vls3}. \\ 
In 
particular, the conformal invariance gives rise to a repulsive interaction of 
the string with the background which precisely cancels the dominant attractive 
term arising from gravity. \\
We perform both semi-classical and canonical string quantization, in order to 
see the effect of the conformal invariance of the background on the string 
mass spectrum. Both approaches yield that the high-mass states are governed 
by 
\begin{displaymath}
m \sim HN (N \in N_0, N ``\mathrm {large}''), 
\end{displaymath}
where {\it m} is the string mass and {\it H} is the Hubble constant. \\
It follows that the level spacing grows proportionally to N :
\begin{displaymath}
\frac {d (m^2 \alpha ^{'})}{dN} \sim N, 
\end{displaymath}
while the string entropy goes like 
\begin{displaymath}
S \sim \sqrt{m}.
\end{displaymath}
Moreover, it follws that there is no Hagedom temperature, so that the partition function is well defined at any positive temperature. \\ 
All results are compared with the analogue results in anti-de Sitter spacetime, which is a nonconformal invariant background.It appears that conformal 
invariance 
{\it simplifies} the mathematics of the problem but the physics remains mainly 
{\it unchanged}. Differences between conformal and non-conformal backgrounds 
only appear in the intermediate region of the string mass spectrum, but these 
differences are minor. For low and high masses, the string mass spectra in 
conformal and non-conformal backgrounds are identical. \\ Interestingly enough, 
conformal invariance fixes the value of the spacetime curvature to be 
$-69/(26\alpha ^{'})$. \\ \\
It has been known for some time that the SL(2,R) WZWN model reduces to 
Liouville theory. In ref \cite{ls6} we give a direct and physical derivation 
of this result based on the classical string equations of motion and the 
proper string size. This allows us to extract precisely the physical effects of the metric and antisymmetric tensor, respectively, on the {\it exact} string 
dynamics in the SL(2,R) background. The general solution to the proper string 
size has been also found \cite{ls6}. \\
We show that the antisymmetric tensor (corresponding to conformal invariance) 
generally gives rise to repulsion, and it precisely cancels the dominant 
attractive term arising from the metric. Both the sinh-Gordon and the 
cosh-Gordon sectors of the string dynamics in non-conformally invariant AdS 
spacetime reduce here to the Liouville equation (with different signs of the 
potential), while the original Liouville sector reduces to the free wave 
equation. \\
Only the very large classical string size is affected by the torsion. Medium 
and small size string behaviors are unchanged. \\
We also find illustrative classes of string solutions in the SL(2,R) background: dynamical closed as well as stationary open spiralling strings, for which 
the effect of torsion is somewhat like the effect of rotation in the metric. 
Similarly, the string solutions in the 2+1 BH-AdS background with torsion and 
angular momentum are fully analyzed \cite{ls6}.

\end{document}